\documentstyle[12pt,amssymb,latexsym,amsmath]{article}
\begin{document}

\title{ Generalized hierarchy of matrix Burgers type and
$n$-wave equations}

\author{ Alexandre I. Zenchuk\\
Center of Nonlinear Studies of\\ L.D.Landau Institute 
for Theoretical Physics  \\
(International Institute of Nonlinear Science)\\
Kosygina 2, Moscow, Russia 119334\\
E-mail: zenchuk@itp.ac.ru\\}
\maketitle

\begin{abstract}
This article concerns  the dressing method
for solving of  multidimensional nonlinear Partial Differential
Equations. In particular,  we join hierarchy of matrix Burgers type 
equation with
hierarchies of equations integrable by the Inverse Spectral Transform
(IST). Example of  resonance interaction
of wave packets  in (3+1)-dimensions
is given.  
\end{abstract}


 \section{Introduction}

It is well known  that  different versions of the
dressing method \cite{ZSh1,ZSh2,ZM,BM,OSTT} are very successful 
tools for solving of so called
completely integrable nonlinear Partial Differential Equations
(PDE). These equations, in turn, have wide application area in
different branches of physics, such as hydrodynamics, plasma
physics, superconductivity, nonlinear optics.  
Although till recently dressing methods have been used only for PDE
integrable by the Inverse Spectral Transform (IST),
it has been shown  in \cite{Z_Sato} 
that there is another type of equations  (maybe not 
integrable by IST) which admit properly modified dressing procedure 
for construction of large manifold of their solutions.
But the technique proposed there  left opened many questions. For instance, 
it is not clear 
whether derived nonlinear PDE can be linearized by some
substitution \cite{C_int1,C_int2}. Also it was difficult to
characterize the manifold  of available
solutions. 

In this paper we 
replace algebraic operator with integral one, generalize system
of equations introducing the set of additional parameters
(independent variables of nonlinear PDE) and modify significantly
the
algorithm given in \cite{Z_Sato}.
This allows us to simplify description  of PDE's properties, 
exhibit more information about 
 solution manifold as well as  relations to the
classical solvable (both linearizable and integrable by IST) 
PDE. 
Although all statements following hereafter  can be proved, we omit most
of the proofs 
for the sake of brevity. They will be given in different paper.

Thus, the  basic object  is the following $N\times N$ matrix 
integral equation 
 \begin{eqnarray}\label{basic}
\Phi\equiv \Phi(\lambda,\mu;t)=\int\limits_{D_\nu} \Psi(\lambda,\nu;t) 
U(\nu,\mu;t) d\nu\equiv \Psi*U,
\end{eqnarray}
where $\lambda=(\lambda_1,\dots,\lambda_{N_\lambda})$,
$\mu=(\mu_1,\dots,\mu_{N_\mu})$, $\nu=(\nu_1,\dots,\nu_{N_\nu})$
are vector variables with  different length in general;
integration is over whole space $D_\nu$ of the appropriate vector parameter;
$\Phi$, $\Psi$ and $U$ are $N\times N$ matrix functions of
arguments.
Star means integration over space of  inner variable: 
$f*g\equiv \int\limits_{D_\nu} f(\lambda_1,\dots,\nu) g(\nu,\dots,\mu_n)d\nu$.
We require that $\Psi$ is invertible operator,
i.e. equation (\ref{basic}) can be solved uniquely for the
function $U$. By definition, operator ${\cal{A}}(\lambda,\mu)$ is
invertible, if there are operators ${\cal{A}}^{-1}_L(\lambda,\mu)$
and ${\cal{A}}^{-1}_R(\lambda,\mu)$
such that
$\int\limits_{D_\nu}{\cal{A}}(\lambda,\nu){\cal{A}}^{-1}_R(\nu,\mu) d\nu =
\int\limits_{D_\nu}{\cal{A}}^{-1}_L(\lambda,\nu){\cal{A}}(\nu,\mu)
d\nu =
\delta(\lambda-\mu)$.
 The functions $\Phi$ and
$\Psi$ are related  by means of the  {\it compatible}
system of linear integral-differential equations, which, 
on the other hand, introduce (infinite) set of   additional parameters
$t=(t_1,t_2\dots)$ 
(independent variables of nonlinear PDE):
\begin{eqnarray}\label{gen}
M_i*\Psi=\sum_{k} L_{ik}*\Phi*C_{ki}, \;\;
C_{ki}=C_{ki}(\lambda,\mu;t),\;\;i=1,2,\dots
\end{eqnarray}
where $M_j=M_j(\lambda,\mu;\partial_{t_1},\partial_{t_2},\dots)$ 
are first order and
 $L_{jk}=L_{jk}(\lambda,\mu;
 \partial_{t_1},\partial_{t_2},\dots)$  are arbitrary order 
 linear differential
operators with matrix coefficients depending on $\lambda$ and
$\mu$.
This overdetermined system together with its compatibility
condition defines $\Psi$ and $\Phi$. 
 Finally, the same compatibility condition
 with substitution $\Phi$ from the eq.(\ref{basic})  results in 
nonlinear PDE whose solution is expressed in terms of $U$. 

After this preliminary discussion we 
 derive some general equations using the following simplified version of 
 the system (\ref{gen}):
\begin{eqnarray}\label{x}
\Psi_{t_i}=S_i*\Phi *C_i,\;\;i=1,2,\dots,
\end{eqnarray}
where $S_i(\lambda,\mu;t)$ and $C_i(\lambda,\mu;t)$ are known 
functions of  $t$, which will be seen later. 
The compatibility condition for the system (\ref{x}) has the form
\begin{eqnarray}\label{comp_Phi}
{S_i}_{t_j}*\Phi*C_i-{S_j}_{t_i}*\Phi*C_j +
{S_i}*\Phi*{C_i}_{t_j}-{S_j}*\Phi*{C_j}_{t_i} +\\\nonumber
{S_i}*{\Phi}_{t_j}*{C_i}-{S_j}*{\Phi}_{t_i} *{C_j}=0,
\end{eqnarray}
which is linear system of compatible integral-differential
equations for the function $\Phi$.
Solving this equation, substituting result in
(\ref{x}) and integrating it, we obtain the expression for
$\Psi$:
$\Psi(\lambda,\mu;t)=\partial^{-1}_{t_1}
(S_1*\Phi*C_1)(\lambda,\mu;t)
+E(\lambda,\mu)+F(\lambda,\mu;t_2,t_3,\dots)$.
Here $E$ is invertible operator,
function $F$ provides compatibility of the system (\ref{x}). 
 Being
invertible,
operator $\Psi$  provides unique solution to the eq.(\ref{basic}).

On the other hand,
 eq.(\ref{comp_Phi}) may be given another form after substitution
eq.(\ref{basic}) for $\Phi$ and (\ref{x}) for $\Psi_{t_i}$
\begin{eqnarray}\label{comp_U}
{S_i}_{t_j}*\Psi*U*C_i-{S_j}_{t_i}*\Psi*U*C_j +
{S_i}*\Psi*U*{C_i}_{t_j}-{S_j}*\Psi*U*{C_j}_{t_i} +\\\nonumber
{S_i}*(S_j*\Psi*U*C_j*U+\Psi*U_{t_j})*{C_i}-
{S_j}*(S_i*\Psi*U*C_i*U+\Psi*U_{t_i})*{C_j}
=0,
\end{eqnarray}
which is nonlocal equation quadratic in $U$. It may result in
nonlinear PDE
 for the dependent variables, expressed
in terms of $U$, $S_i$ and $C_i$.
To provide this possibility we must impose
specific dependence of the functions $S_i$ and $C_i$ on their
arguments, for instance, in accordance 
to the following set of relations:
\begin{eqnarray}\label{ad1}
S_i(\lambda,\mu;t)=S(\lambda,\mu),\;\; 
\Phi=S*\Phi + \chi, \;\; \chi=\chi(\lambda,\mu;t)
,\\\label{ad2}
 C_i(\lambda,\mu;t)=
 \int\limits_{D_\nu} A_i(\lambda,\nu) p_1(\nu;t) p_2(\mu) d\nu+
c_1(\lambda) B_i c_2(\mu;t)\equiv A_i*p_1(t) p_2 + c_1 B_i c_2(t),\\\label{ad3}
A_i*c_1=c_1 B_i,\\\nonumber
\end{eqnarray}
where $A$ is invertible
 operator, $A_i*A_j=A_j*A_i$, $[B_i,B_j]=0$ and
 $[*,*]$ means commutator of two matrices.
Eqs.(\ref{ad1}-\ref{ad3}) split
 eq.(\ref{comp_Phi})  into the following set of three
integral-differential equations for $\Phi$, $p_1$ and $c_2$ :
\begin{eqnarray}\label{BN1}
S*\Phi_{t_j}* A_1 - S*\Phi_{t_1} *A_j=0,\\\label{c1}
A_1 *({{p_1}})_{t_j}- A_j *({{p_1}})_{t_1}=0,\\\label{c2}
B_1 {c_2}_{t_j}-B_j {c_2}_{t_1}=0.
\end{eqnarray}
Then one has the following nonlinear equation instead of (\ref{comp_U})
\begin{eqnarray}\label{comp_U2}
\Psi*(U_{t_j}*A_1-U_{t_1}*A_j
+U*C_j*U*A_1-U*C_1*U*A_j)+&&\\\nonumber
\chi_{t_1}*A_j-\chi_{t_j}*A_1 + \chi*(C_1*U*A_j-C_j*U*A_1)&=&0.
\end{eqnarray}
Note, that reduction leading to the equations introduced in 
\cite{Z_Sato} will be discussed in different paper.
Here
we consider two other examples   
of multidimensional systems. First of them (Sec.2.) represents
combination of linearizable  (Burgers type) and completely integrable
($n$-wave) $(3+1)$-dimensional 
systems,
having solutions depending on arbitrary functions of three
variables. Second example (Sec.3) is another generalization
of the matrix $n$-wave system
\cite{T}. Properly introduced multiple scales expansion 
of this system results in the
multidimensional ((3+1)-dimensional in our case)
equation describing resonance interaction of wave
packets. Its solutions may depend on arbitrary functions of two
variables.  Both examples have extension into $(n+1)$-dimensions
with arbitrary $n$.

\section{Generalized hierarchy  of linearizable (Burgers type)  
and  integrable by IST ($n$-wave) systems}

In this section $S(\lambda,\mu)=\delta(\lambda-\mu)$, $\chi=0$,
$A_j=\underbrace{A*\dots*A}_{j}\equiv A^j$, $B_j=B^j$,
 where $A(\lambda,\mu)$ is invertible operator and
 $B$ is nondegenerate
constant matrix. Thus $\Psi_{t_i}=\Phi*C_i$ with $C_i$ given by
(\ref{ad2}).
 After applying operator $\Psi^{-1}$ from
the left to the eq.(\ref{comp_U2}) one results in
\begin{eqnarray}
E_j=U_{t_j}*A +U*A^j*p_1  p_2*U*A+U*c_1 B^j c_2*U*A
-&&\\\nonumber
(U_{t_1}*A^j +U*A*p_1 
p_2*U*A^j+
U*c_1 B c_2*U*A^j)&=&0
\end{eqnarray}
We may derive nonlinear system for the functions
 \begin{eqnarray}\label{pot}
 u=p_{2}*U*c_1,\;\;
 q_n=p_2*U*A^n* p_1,\;\;
 v_n=\partial^n c_2*U* c_1,\;\;
 w_{nm}=\partial^n c_2*U* A^m *p_1,
 \end{eqnarray}
 which has the following "short" form
 \begin{eqnarray}
p_{2}*E_j*c_1 = 0,\;\;
 p_2*E_j*A^n* p_1 = 0,\;\;
 \partial^n c_2*E_j* c_1 = 0,\;\;
 \partial^n c_2*E_j* A^m *p_1 = 0,
 \end{eqnarray}
 or, extended form
 \begin{eqnarray}\label{Burgers_n1}
 u_{t_j} - u_{t_1} B^{j-1} + q_j u - q_1 u B^{j-1} +u B^j v_0 - 
 u B v_0 B^{j-1} &=&0,\\
 \nonumber
 {q_n}_{t_j}-{q_{n+j-1}}_{t_1} +q_j q_{n}-q_1
 q_{n+j-1} + u B^j w_{0n}   -u B
 w_{0(n+j-1)}&=&0,\\\label{Burgers_n2}
 {v_n}_{t_j}- {v_n}_{t_1} B^{j-1}+w_{nj}u-w_{n1} u B^{j-1} -
  B^{j-1} v_{n+1} + v_{n+1} B^{j-1} +\\\nonumber
   v_n B^j v_0 - v_n
B v_0 B^{j-1}&=&0,\\\label{Burgers_n3}
 {w_{mn}}_{t_j} - {w_{m(n+j-1)}}_{t_1}  +
  w_{mj} q_n -  w_{m1} q_{n+j-1}- B^{j-1} w_{(m+1)n}
  +&&
  \\\label{Burgers_n4}
 w_{(m+1)(n+j-1)} +v_m B^j w_{0n} -
  v_m B w_{0(n+j-1)}&=&0.
  \end{eqnarray}
  The complete system of pure PDE is represented by the
  following set: eq. (\ref{Burgers_n1}) with $j=2$,   
 eqs. (\ref{Burgers_n2}) and (\ref{Burgers_n3}) with $j=2,\;3$, 
  eq. (\ref{Burgers_n4}) with $j=2,\;3,\;4$. Thus, this system is 
  (3+1)-dimensional.
 It  
 may be given the compact form if one introduces
  column of matrices $u,q_n,v_n,w_{nm}$:
  $\chi=[u,q_1,q_2,\dots, v_1,v_2,\dots,
   w_{00},w_{10},w_{01},\dots]^T$:
   \begin{eqnarray}\label{short}
   \sum_{l=1}^4 \sum_{mn}V_{lijmn}\partial_{t_l} \chi_{mn}
   +\sum_{klmn}T_{ijklmn}\chi_{kl}\chi_{mn}=0,
   \end{eqnarray}
   where $V_{lijmn}$ and $T_{ijklmn}$ are constants expressed in
   terms of the elements of the matrix $B$.

  Physical application of the eqs. (\ref{Burgers_n1}-\ref{Burgers_n4}) 
  is not found yet. 
In particular, it reduces into the following (2+1)-dimensional
systems:
\begin{enumerate}
\item 
{\it Matrix Burgers type system} (i.e. linearizable)  for the function
$q_{0}$, if
$c_1=0$ or $N=1$. 
\item 
{\it Matrix $n$-wave equation} ($n=N(N-1)/2$)  for the function
$v_0$,
if
$p_1=0$.
\end{enumerate}


\subsection{Construction of solutions}

First, one needs to solve the system
(\ref{ad3}-\ref{c1}) for the functions $c_1$, $\Phi$, $p_1$,
$c_2$:
\begin{eqnarray}
A*c_1&=&c_1 B,\\
\Phi(\lambda,\mu;t)&=&\int\limits_{\Omega_k}\int\limits_{D_\nu}
\Phi_0(\lambda,\nu;k)e^{\eta_1(\nu;k,t)}
\phi_0(\nu,\mu;k)dk d\nu,\;\;\\
p_1(\lambda;t)&=&\int\limits_{\Omega_k}\int\limits_{D_\nu}
p_0(\lambda,\nu;k) e^{\eta_2(\nu;k,t)}p_{10}(\nu;k)dk d\nu,\;\;
 \\
 c_2(\lambda;t)&=&\int\limits_{\Omega_k} 
 e^{k\sum_i B^i t_i} c_{20}(\lambda;k ) dk
\end{eqnarray}
where
$\eta_i(\mu;k,t)=\sum_{j=1}^4 \eta_{ij}(\mu;k)
 t_j$, $i=1,2$,
 $[\eta_{ij},\eta_{ik}]=0$, $\det(\eta_{ij})\neq 0$.
  Parameter $k$ is complex in general, integration is over
whole complex plane $\Omega_k$ of this parameter.
Function $c_{20}$ is arbitrary and functions $\phi_0$ and $p_{0}$ 
solve the following system:
\begin{eqnarray}\label{phi0}
\eta_{1j}(\nu;k)\phi_0(\nu,\mu;k)&=&
\int\limits_{D_{\nu_1}}\eta_{1(j-1)}(\nu;k)
\phi_0(\nu,\nu_1;k) A(\nu_1,\mu) d\nu_1, 
\\\label{p0}
p_0(\lambda,\nu;k)\eta_{2j}(\nu;k)&=&
\int\limits_{D_{\nu_1}} A(\lambda,\nu_1) p_0(\nu_1,\nu;k)\eta_{2(j-1)}(\nu;k)d\nu_1,
\;\;j=2,3,\dots. 
\end{eqnarray}
Functions $\eta_{j1}$ ($j=1,2$) are arbitrary, 
while $\eta_{jn}$ with $n>1$ provide
compatibility of eqs. (\ref{phi0}) and (\ref{p0}).

Now one can integrate (\ref{x})   to get $\Psi$ ($j=1$):
\begin{eqnarray}
\Psi(t)&=&E+
\partial_{t_1}^{-1}\left[\Phi(t)*A*p_1(t)
p_2+\Phi(t)*c_1 c_2(t)\right],
\end{eqnarray}
where $E=E(\lambda,\mu)$ is invertible operator independent on
$t$. For instance, $E(\lambda,\mu)=\delta(\lambda-\mu)$.
Next, find $U$ from (\ref{basic}): $ U=\Psi^{-1}*\Phi$. In
general, operator $\Psi^{-1}$ 
 can be constructed  only numerically,
unless 
$\Phi_0$ is
degenerate ($\Phi_0(\lambda,\mu;k)=
\sum_n \Phi_{01}(\lambda) \Phi_{02}(\mu;k) $) and explicit form
for $E^{-1}_L$ is known. In this case $\Psi^{-1}$ may be found
analytically, following the procedure proposed, for instance,  in
\cite{K},
where $\bar\partial$-problem with degenerate kernel has been solved.
Similarly, eqs. (\ref{phi0}) and (\ref{p0})
can be solved numerically, unless $A$ has the following
structure:
$A(\lambda,\mu)= A_0(\lambda,\mu) + \sum_{j} A_{j1}(\lambda)
A_{j2}(\mu)$, where operator $A_0$ is invertible with known
analytical form for  ${A_0}_R^{-1}$.  For instance,
$A_0(\lambda,\mu)=\delta(\lambda-\mu)$.
Then
compatibility condition of the system (\ref{phi0}) and (\ref{p0})
produces dispersion relations in the form
$\eta_{1n} (\mu;k)=\eta_{11}(\mu;k)\; F_1[(\phi_0*A_{i1})(\mu;k),
\;i=1,2,\dots]$,
$\eta_{2n}(\mu;k)= 
F_2[(A_{i2}*p_0)(\mu;k), \;i=1,2,\dots]\;
\eta_{21}(\mu;k)$,
 where $F_i$ are given matrix functions of matrix  arguments.

Finally, one can show that solutions of our (3+1)-dimensional system 
(\ref{Burgers_n1}-\ref{Burgers_n4}) constructed in accordance
with definitions (\ref{pot}) may
 depend on arbitrary functions of
three  real  parameters, for instance, $t_1$, $t_2$ and $t_3$.
This is owing to the factor $\Psi*A*p_1$.

\section{Resonance  wave interaction   in 
(3+1)-dimensions}

In this section we consider eqs.(\ref{ad1}-\ref{c2}) with
 $S(\lambda,\mu)\neq \delta(\lambda-\mu)$, $\chi\neq 0$,  $
 p_1=0$ and $\Psi_{t_i}=S*\Phi*C_i$.
It is convenient to apply operator $c_1$
to the eqs.(\ref{basic})  and (\ref{BN1}) 
 from the right, giving them the form:
\begin{eqnarray}\label{new_basic}
S*\tilde\Phi+\tilde \chi&=&\Psi*\tilde U,\;\;\tilde U=U*c_1,\;\;
\tilde \chi=\chi*c_1,\\
\label{ex_comp}
 (S*\tilde \Phi)_{t_j}&=&(S*\tilde \Phi)_{t_1} B_j,\;\;B_1=I,
 \end{eqnarray}
 $I$ is identity matrix, $B_i$ are diagonal matrices  and  $N\ge
 4$.
 Let $\tilde\chi_{t_j}(\lambda;t)=\tilde \chi(\lambda ;t) a_j$,
 where
 $a_j$ are constant matrices.
 We will need the following notations: $b_j=a_1 B_j-a_j$,
 $V_0=c_2*\tilde U$ and $V_1= {c_2}_{t_1}*\tilde U$.
 Nonlinear eq.(\ref{comp_U2}) gets the following form after
 applying operator $c_1$ from the right:
 \begin{eqnarray}\label{nwave_lin}
 \Psi*(\partial_{t_j}\tilde U - \partial_{t_1}\tilde U B_j +
 \tilde U [B_j ,V_0] )+\tilde \chi (b_j-[B_j, V_0])=0.
\end{eqnarray}
 Now assume that ${\mbox{det}} (b_j-[B_j,V_0])\neq 0$ for all $j$
 and use two equations (\ref{nwave_lin}) with indexes $j$ and $k$, $j\neq k$
 to  eliminate function
 $\tilde \chi$. After applying   
 operator $c_2*\Psi^{-1}$ from the left to the resulting
 equation, we receive:
\begin{eqnarray}\label{nwave0}
 (\partial_{t_k}V_0 - \partial_{t_1}V_0  B_k  +
[V_1, B_k]+V_0 [B_k ,V_0] )(b_k-[B_k, V_0])^{-1}&=&\\\nonumber
(\partial_{t_j} V_0 - \partial_{t_1} V_0 B_j +
[V_1, B_j]+V_0 [B_j ,V_0])(b_j-[B_j, V_0])^{-1}.&&
\end{eqnarray}
Next, let us introduce different scales for
variables $t_k$, $V_0$, $V_1$:
$\partial_{t_k} \to \epsilon \partial_{t_k}$,
$V_0= \epsilon v$, $V_1=\epsilon^2 v_1$.
Keeping only leading terms, we get from the  eq.(\ref{nwave0}):
\begin{eqnarray}\label{nwave2}
E_{k}\equiv v_{t_1} (B_j b_j^{-1}-B_k b_k^{-1} )+ v_{t_k}
b_k^{-1}- v_{t_j}
b_j^{-1} +\\\nonumber
[v_1,B_k]b_k^{-1}-[v_1,B_j]b_j^{-1}+
v[v,B_j]b_j^{-1} -v[v,B_k]b_k^{-1}=0.
\end{eqnarray}
Thus the complete system is represented by the pair of equations
(\ref{nwave2}), $E_{k}$ and $E_{n}$,  $k\neq n$. 
One can see that the following combination of these equations
has no function $v_1$ and contains only off-diagonal elements of
$v$:
\begin{eqnarray}
E_k \left(B_n b_n^{-1} - B_j b_j^{-1}\right)-
E_n \left(B_k b_k^{-1} - B_j b_j^{-1}\right)+&&\\\nonumber
B_j (E_k-E_n) b_j^{-1} - B_n E_k b_n^{-1} +  
B_k E_n b_k^{-1}&=&0.
\end{eqnarray}
Let $j=2$, $k=3$, $n=4$ and write  this equation 
in the following form
\begin{eqnarray}\label{nwave3}
 \sum_{n=1}^4 s_{nij} \partial_{t_n} v_{ij} + 
\sum_{k:k\neq i\neq j} T_{ikj} v_{ik} v_{kj}=0,\;\;i\neq j,
\end{eqnarray}
where $s_{kij}$,  and $T_{ikj}$ are constants,
expressed in terms of the elements of the  matrices $B_j$ and $b_j$.
If $v_{ij}$ are real, then this equation describes  resonance  
interaction of wave packets.

Reduction $t_k \to i t_k$,  $v_{ij}=\bar v_{ji}$, with real
$s_{nij}$ and $T_{ikj}$ , $s_{nij}=s_{nji}$, $T_{ikj}=T_{jki}$ (bar means
complex conjugated value) transforms the (3+1)-dimensional eq.(\ref{nwave3})
  into (2+1)-dimensional $n$-wave
equation with 
independent variables
$\tau_k=t_k+t_1$, $k=2,3,4$.

\subsection{Construction of  solutions}

In this section we give the algorithm for construction the
solution $V_0$ to the 
 eq.(\ref{nwave0}).

Solutions of the eq.(\ref{ex_comp}) and expression for
$\tilde \chi$ have the form:
\begin{eqnarray}\label{sol1}
S*\tilde\Phi(\lambda)=\int\limits_{\Omega_k} 
\tilde\Phi_0(\lambda,k) e^{k\sum_n B_n t_n}
dk,\;\;
c_2(\lambda)=\int\limits_{\Omega_k} e^{k\sum_n B_n t_n } c_{20}(\lambda,k)  dk,\\
\tilde\chi(\lambda)= \chi_0(\lambda) e^{\sum_n a_n t_n}.
\end{eqnarray}
To find $\Psi$ we integrate eq.(\ref{x}) ($j=1$, remember that
$B_1=I$):
\begin{eqnarray}
\Psi(\lambda,\mu)=
\int\limits_{\Omega_k}\int\limits_{\Omega_q} \tilde\Phi_0(\lambda,k) 
e^{(k+q)\sum_n B_n t_n } 
  c_{20}(\mu,q) \frac{ d k d q}{k+q}+ \delta(\lambda-\mu).
\end{eqnarray}
Thus
\begin{eqnarray}
\tilde U(\lambda)=S*\tilde \Phi(\lambda)-
\int\limits_{\Omega_k}\int\limits_{\Omega_q} \tilde\Phi_0(\lambda,k) 
e^{\sum_n B_n t_n (k+q)} 
  \phi(q) \frac{ d k d q}{k+q} + \tilde \chi(\lambda),\;\;
  \phi=c_{20}* \tilde U
\end{eqnarray}
and
\begin{eqnarray}\label{solf}
V_0=
c_2*\tilde U .
\end{eqnarray}
Unknown function $\phi$, related with $\tilde U$, can be found
only numerically in general case, unless functions
$\tilde\Phi(\lambda,k)$ is degenerate \cite{K}.
Eq.(\ref{solf}) shows that 
 $V_0$ may depend on $N\times N$ matrix
function of two real variables, for instance, $t_1$ and $t_2$.

Regarding the multi-scale expansion given by
eqs.(\ref{nwave2}), one should 
replace $t_n\to \epsilon
t_n$, in formulae (\ref{sol1}-\ref{solf}) and take arbitrary
functions  $\tilde \Phi_0$  and $\chi_{0}$ 
proportional to ${\epsilon}$. Thus $V_0\sim \epsilon$.

\section{Conclusions}

Working with dressing methods we underline two directions:
(a) increase of dimension of  solvable nonlinear PDE and (b)
provide rich class of their solutions. Nonlinear PDE derived with
our algorithm admit infinite set of commuting flows
corresponding to different parameters $t_j$. 
Since general equations are ruther complicated (see
 (\ref{Burgers_n1}-\ref{Burgers_n4}), (\ref{nwave0})), 
 the reasonable problem is construction of their reductions,
 which would exhibit physical application of these systems. 
 Another way is multi-scale expansion of general systems, 
 which in our case reveals 
(3+1)-dimensional equation describing resonance interaction of wave
packets (see eq.(\ref{nwave2}) and (\ref{nwave3})). 

Author thanks Prof. S.V.Manakov for discussion.
This work was supported by 
RFBR grants 03-01-06122 and 1716.2003.1.

\end{document}